\documentclass{aa}
\usepackage[varg]{txfonts}
\usepackage{natbib}
\usepackage{graphicx}
\graphicspath{{figures/}}
\usepackage{hyperref}
\hypersetup{
    colorlinks=false, 
    linkcolor=blue, 
    urlcolor=red, 
    linktoc=all 
}
\bibpunct{(}{)}{;}{a}{}{,}
\usepackage{xcolor}
\usepackage{multirow}

\newcommand{\Halpha}{\ifmmode {\rm H}\alpha \else H$\alpha$\fi}
\newcommand{\Hbeta}{\ifmmode {\rm H}\beta \else H$\beta$\fi}
\newcommand{\Hgamma}{\ifmmode {\rm H}\gamma \else H$\gamma$\fi}
\newcommand{\Hdelta}{\ifmmode {\rm H}\delta \else H$\delta$\fi}
\newcommand{\Lya}{\ifmmode {\rm Ly}\alpha \else Ly$\alpha$\fi}
\newcommand{\Lyb}{\ifmmode {\rm Ly}\beta \else Ly$\beta$\fi}
\newcommand{\HeI}{\ifmmode {\rm He}\,\textsc{i}\,\lambda5876 \else 
                  He\,\textsc{i}\,$\lambda5876$\fi}
\newcommand{\HeII}{\ifmmode {\rm He}\,\textsc{ii}\,\lambda4686 \else 
                   He\,\textsc{ii}\,$\lambda4686$\fi}

\newcommand{\ciii}{\ifmmode {\rm C}\,\textsc{iii}] \else C\,\textsc{iii}]\fi}
\newcommand{\civ}{\ifmmode {\rm C}\,\textsc{iv} \else C\,\textsc{iv}\fi}

\newcommand{\eddrat}{\ifmmode{\lambda_{\text{Edd}}} \else  $\lambda_{\text{Edd}}$ \fi}
\newcommand{\mbh}{\ifmmode{M_{\text{BH}}} \else  $M_{\text{BH}}$ \fi}
\newcommand{\athena}{\textsl{NewAthena}}
\newcommand{\prima}{\textsl{PRIMA}}
\newcommand{\nh}{N_{\text{H}}}
\newcommand{\lx}{L_{\text{{X}}}}
\newcommand{\lbolo}{L_{\text{{bolo}}}}
\newcommand{\lognh}{\log{(N_{\text{H}}/\rm{cm^{-2}})}}
\newcommand{\Nh}{\log{(N_{\text{H}}/\rm{cm^{-2}})}}
\newcommand{\loglx}{\log{(L_{\text{{X}}}/\rm{erg\,s^{-1}})}}
\newcommand{\Lx}{\log{(L_{\text{{X}}}/\rm{erg\,s^{-1}})}}
\newcommand{\Lbolo}{\log{(L_{\text{{bolo}}}/\rm{erg\,s^{-1}})}}
\newcommand{\Lbolosun}{\log{(L_{\text{{bolo}}}/\rm{L_{\odot}})}}
\begin{document}

\title{A New Hope for Obscured AGN: The PRIMA-NewAthena Alliance}

\author{Luigi Barchiesi\thanks{\email{luigi.barchiesi@uct.ac.za}}\inst{1,2,3}
  \and F.~J.~Carrera\inst{4}
    \and C.~Vignali\inst{5,6}
        \and F.~Pozzi\inst{5,6}
        \and L.~Marchetti\inst{1,2,3}
        \and C.~Gruppioni\inst{6}
        \and I.~Delvecchio\inst{6}
        \and L.~Bisigello\inst{7}
        \and F.~Calura\inst{6}
        \and J.~Aird\inst{8}
        \and M.~Vaccari\inst{1,2,3,9}}


\institute{Department of Astronomy, University of Cape Town, 7701 Rondebosch, Cape Town, South Africa
  \and Inter-University Institute for Data Intensive Astronomy, Department of Astronomy, University of Cape Town, 7701 Rondebosch, Cape Town, South Africa
  \and INAF–Istituto di Radioastronomia, Via Piero Gobetti 101, I-40129 Bologna, Italy
  \and IFCA (CSIC-University of Cantabria), Avenida de los Castros, 39005, Santander, Spain
  \and Dipartimento di Fisica e Astronomia (DIFA) ``Augusto Righi'', Universit\`a di Bologna, Via Gobetti 93/2, 40129, Bologna, Italy
  \and Istituto Nazionale di Astrofisica (INAF) – Osservatorio di Astrofisica e Scienza dello Spazio (OAS), Via Gobetti 101, 40129, Bologna, Italy
  \and Istituto Nazionale di Astrofisica (INAF) - Osservatorio Astronomico di Padova, Via dell'Osservatorio 5, 35122 Padova, Italy
  \and Institute for Astronomy, University of Edinburgh, Royal Observatory, Edinburgh EH9 3HJ, UK
  \and Inter-University Institute for Data Intensive Astronomy, Department of Physics and Astronomy, University of the Western Cape, 7535 Bellville, Cape Town, South Africa}


\date{Accepted for publication on JATIS, 21 March 2024 }

\abstract{Understanding the AGN-galaxy co-evolution, feedback processes, and the evolution of Black Hole Accretion rate Density (BHAD) requires accurately estimating the contribution of obscured Active Galactic Nuclei (AGN). However, detecting these sources is challenging due to significant extinction at the wavelengths typically used to trace their emission. We evaluate the capabilities of the proposed far-infrared observatory \prima\ and its synergies with the X-ray observatory \athena\ in detecting AGN and in measuring the BHAD. Starting from X-ray background synthesis models, we simulate the performance of \athena\ and of \prima\ in Deep and Wide surveys. Our results show that the combination of these facilities is a powerful tool for selecting and characterising all types of AGN. While \athena\ is particularly effective at detecting the most luminous, the unobscured, and the moderately obscured AGN,  \prima\ excels at identifying heavily obscured sources, including Compton-thick AGN (of which we expect 7500 detections per $\rm{deg^2}$). We find that \prima\ will detect $\sim60$ times more sources than \textsl{Herschel} over the same area and will allow us to accurately measure the BHAD evolution up to $z\sim8$, better than any current IR or X-ray survey, finally revealing the true contribution of Compton-thick AGN to the BHAD evolution.
}

\maketitle 

\section{Introduction}
\label{sect:intro}  
In the mid-1990s, the discovery of a tight correlation between the mass of supermassive black holes (SMBHs) and the stellar mass of the host galaxies\citep{kormendy95,magorrian98,ferrarese00,gebhardt00,kormendy13} revealed a mutual influence between SMBHs and their host galaxies during their evolution. Notably, the relation between SMBH mass and stellar mass, as well as with bulge mass, and gas and stellar velocity dispersions suggests a link between SMBH growth and star formation (SF), leading to the formulation of the AGN-galaxy co-evolution paradigm \citep{hopkins08,lapi14,lapi18}. In this scenario, an intense phase of SF is triggered by wet mergers, particularly in the most luminous and massive systems \citep{silk98,dimatteo05,treister12,lamastra13}, or by in-situ processes, such as rapid gas inflow and cooling of gas clumps \citep{lapi18}. A fraction of the galaxy’s gas reservoir is funneled toward the SMBH, triggering active galactic nucleus (AGN) activity. Thus, this phase is marked by the concurrent growth of the SMBH and the galaxy stellar mass. Due to the large amounts of gas, most of the AGN radiation is absorbed through photoelectric absorption, causing the source to appear as an obscured AGN (with column density $\nh \geq 10^{22}\,\rm{cm^{-2}}$). This phase is likely associated with the growth of obscured AGN in strongly star-forming submillimeter galaxies \citep{archibald02,almaini03,alexander05}.\par
However, we are far from having a complete picture of the galaxy co-evolution, and several key questions remain open: What mechanisms drive the co-evolution? What are the timescales involved? Is the co-evolution externally triggered (e.g., by galaxy mergers) or intrinsic to the galaxy evolution? To answer these questions, we need a comprehensive selection and study of obscured AGN across cosmic times.\par 
The connection between the host galaxy and the SMBH is also evident in the similar evolutions of the BH Accretion rate Density (BHAD) and the Star Formation Rate Density (SFRD). Both peak at the ``cosmic noon'' ($z\sim1-2$) and are characterized by significant uncertainties at redshifts $z > 3$\citep{madau14, heckman14, vito18}, primarily due to the challenges in detecting and accurately quantifying the contribution of obscured (i.e., dust-rich) sources at high redshifts. Furthermore, while the co-evolution up to the ``cosmic noon'' is widely accepted, there is no consensus yet on the fact that the SFRD and the BHAD still show similar evolution at higher redshift. In particular, while some works find a $z>3$ decline of the BHAD similar to the one of the SFRD\citep{vito14,aird15,vito18,pouliasis24}, others argue for a flatter evolution \citep{runburg22}. Recently, the uncertainty on the high-$z$ evolution of the BHAD has been exacerbated by the JWST discovery of ``Little Red Dots'' (LRDs), which if we assume to be dust-reddened broad-line AGN, provide BHAD measures more than one magnitude higher than the previous ones\citep{yang23}.\par
A major challenge in addressing these questions is that most SMBH and host galaxy mass growth is expected to occur under heavily obscured conditions, making identifying and studying sources in this phase a daunting task \citep{rowan-robinson97,hughes98,martinez-sansigre05}. In fact, the large quantity of gas and dust fueling both processes absorb the energy emitted by stars (at optical and UV wavelengths) and accreting SMBHs (in the X-ray, UV, and optical regimes), re-emitting it at longer wavelengths, primarily in the infrared (IR, $\sim 1-1000\mu\rm{m}$). One solution is to study the primary emission indirectly, by measuring dust-reprocessed radiation in the IR. The efficacy of identifying obscured AGN by selecting bright mid-IR sources with faint optical or near-IR emission has been demonstrated by \textit{Spitzer} \citep{houck05, weedman06, polletta08} and \textit{WISE} \citep{mateos12, assef13}. Observations in this wavelength range provide insights into the physical processes occurring in obscured regions, allowing estimates of SFRD and BHAD, provided the two contributions can be properly disentangled. Space-based IR observatories enable direct measurement of dust-obscured SF activity, without requiring correction for dust attenuation. Mid- to far-IR photometric observations in deep fields \citep{delvecchio14, schreiber15} have produced estimates of SFRD and BHAD up to redshifts of $z \sim 3$ \citep{gruppioni13, magnelli13, delvecchio14}, using \textsl{Herschel PACS} data \citep[100 and $160\,\mu$m,][]{poglitsch10}. However, the deepest cosmological surveys performed by \textsl{Herschel} at high redshifts have primarily detected the most luminous galaxies \citep[$L_{\rm IR} > 10^{12}\, \rm{L{\odot}}$ at $z \geq 3$,][]{gruppioni13}, and they face challenges in accurately separating AGN and SF contributions, as the \textit{Herschel} mission did not sample the mid-IR part where the AGN emission dominates\citep{pozzi12}. Even when spectral energy distribution (SED) fitting is used to distinguish between these components, the assumption made in modelling them provides an additional source of uncertainties. Consequently, the initial phase of BH-galaxy co-evolution remains elusive and difficult to track.
\par
One of the most effective methods for selecting unobscured or mildly obscured AGN is through X-ray observations, as the radiation from the innermost regions of the AGN can be directly detected \citep{vignali14}. However, the effectiveness of this approach diminishes with increasing column density (at $N_\text{H} \ge 10^{24}\,\text{cm}^{-2}$, the soft X-ray continuum is significantly attenuated). Even in deep field surveys, only a fraction of the most obscured AGN has been uncovered in the X-ray band \citep{tozzi06, lanzuisi13, marchesi16b, delmoro17}.
\par
Summarising, the X-ray band is a powerful tool to select and characterise unobscured AGN, but it fails when heavily obscured sources are involved. On the other hand, the IR band does not suffer from source obscuration, but it has been limited to lower redshift than the X-rays and it relies on being able to properly disentangle the AGN and host-galaxy components. \par
Since the decommissioning of the \textit{Spitzer} and \textit{Herschel} space telescopes, the mid- and far-IR regimes have lacked instruments capable of detecting the emission from obscured galaxies and AGN beyond the local universe (while \textit{WISE} is still operational, it has a wavelength coverage similar to JWST, i.e. it is not able to trace the mid- and far-IR at high-$z$).\par
In this paper, we introduce the key role of the PRobe far-Infrared Mission for Astrophysics (\prima; PI: J. Glenn)\citep{glenn23,bradford22,moullet23} on characterising the population of obscured AGN and in revealing the evolution of the BHAD at high redshift. \prima\ is a cryogenically cooled FIR observatory with a 1.8-meter diameter, currently in the mission concept phase, with potential approval by 2026 and launch in 2032. While the selection and characterisation of AGN and galaxies with  \prima\ is already been investigated in several works \citep{bethermin24,bisigello24} \citep[we refer to the \prima\ GO science book for an extensive list of \prima\ scientific cases][]{moullet23}, here we explore the synergies between \prima\ and the upcoming ESA X-ray observatory New Advanced Telescope for High ENergy Astrophysics (\athena), expected to be launched in the late 2030s, in detecting and characterising obscured AGN at redshifts never reached before.\par
In Section~\ref{sec:instruments} we give a brief description of the \prima\ and \athena\ instruments and surveys. The simulations of the intrinsic AGN predictions and our methods to estimate the detection capabilities of \athena\ and \prima\ are reported in Section~\ref{sec:methods}. Section~\ref{sec:methods_bhad} explains how we simulated \prima\ measurements of the BHAD. The results are shown in Section~\ref{sec:results}, with the discussion of the capabilities of both instruments and of their synergies in Sec~\ref{sec:discussions}. In Section~\ref{sec:conclusions}, we present our conclusions. \par 
Throughout this paper, we adopt the following cosmological parameters: H$_0 = 70\, \text{km}\, \text{s}^{-1}\, \text{Mpc}^{-1}$, $\Omega_{\text{M}} = 0.3$ and $\Omega_{\Lambda} = 0.7$ \citep{spergel03}.

\section{Instruments and surveys}\label{sec:instruments}

\subsection{PRIMA}\label{sec:ins_prima}

\prima\ is a proposed far-infrared (FIR) observatory equipped with a cryogenically cooled 1.8-meter diameter telescope, specifically engineered for ultra-high sensitivity imaging and spectroscopic studies in the $24-235\,\mu$m wavelength range. Its design is optimized for the efficient survey of large areas, making it a powerful tool for wide-field astrophysical observations.
The current \prima\ design incorporates two scientific instruments: \textsl{FIRESS} and \textsl{PRIMAger}. \textsl{FIRESS} is a versatile, multi-mode survey spectrometer covering wavelengths between 24 and 235 µm, offering both low-resolution ($\textrm{R}\sim100$) and high-resolution ($\textrm{R}\sim4400-12000$) in Fourier transform spectrometer mode. \textsl{PRIMAger}, on the other hand, will have two cameras able to operate at the same time: the PRIMA Hyperspectral Imager (PHI)- providing hyperspectral linear variable filters across two bands at $\textrm{R} = 10$ (PHI1 and PHI2) from 24 to 80$\,\mu$m,- and the PRIMA Polarimetric Imager (PPI), that will use 4 filters to cover the $80-261\,\mu$m range with imaging and polarimetric capabilities. Thanks to its cryogenic cooling and advanced kinetic inductance detectors \citep{day03,baselmans12,day24}, \prima\ will provide a mapping speed up to four orders of magnitude higher than \textit{Herschel} at $100\mu$m and at $\sim2$ dex better than \textit{Spitzer} at $24\mu$m\citep{moullet23}. Moreover, it has been shown that the contiguous coverage from $24\,\mu$m to $261\,\mu$m (paired with deblending techniques and possibly shorter wavelengths priors) will allow us to reliably measure source fluxes up to one order of magnitude below the classical confusion limit\citep{donnellan24} \citep[the classical confusion limit being defined as 5 times the confusion noise, the latter obtained from $5\sigma$-clipping the simulated maps until the convergence of the standard deviation][]{bethermin24}.\par
In this work, we will focus only on the \textsl{PRIMAger} instrument, as it is the best suited to perform large area survey. To simulate the \textsl{PRIMAger} Hyperspectral Imager, we approximated it as a series of 12 narrow filters, with center wavelengths and widths closely resembling the $\textrm{R}=10$ capabilities of \textsl{PRIMAger} PHI. For the PRIMA Polarimetric Imager, we focused only on its total intensity sensitivity.  \par
As \prima\ is still in the design phase and an official survey strategy has not been developed yet, we will use two example surveys. A Deep survey covering $1\,\rm{deg}^2$ for a total 1000hr (overhead included), and a Wide survey of 1000hr covering $28\,\rm{deg}^2$. Table~\ref{tab:survey} summarises the survey strategy used in this work. In terms of confusion effects, our current survey strategy reaches sensitivity below the confusion limit starting from the $8^{th}$ filter (PHI2 at $\sim50\mu$m)\citep{donnellan24}. However, it has been demonstrated that even with a $1500\,$hr survey, catalogues with a $95\%$ purity can be produced for the first six filters (i.e., covering the lowest wavelength intervals). By using these as priors, the flux in all the confused bands can be recovered with an accuracy of $20\%$ \citep{donnellan24}.

\begin{table*}
\centering
\caption[Reference surveys used in this work.]{Reference surveys used in this work. t$_{\rm{field}}$ refers to the on-source time per pointings excluding overheads, t$_{\rm{tot}}$ to the survey total time including overheads. Due to its instrument design, \textsl{PRIMAger} is not able to perform snapshots, instead, it relies on panning along one direction when scanning its field. The deepest layer of \athena\ surveys comprises $30$ $300\,$ks-fields for a total of $12\rm{deg^2}$ and $2,500$hr (excluding overhead); however, for our Deep survey, we focused only on $1\rm{deg^2}$ of it to be able to perform comparison with the \prima\ deepest layer. The sensitivities of \prima\ are reported for the most and least sensitive bands (i.e., $\sim25\,\mu$m and $\sim235\,\mu$m, respectively) for a $5\sigma$ detection and have been calculated from the latest \textsl{PRIMAger} characteristics available in \url{https://prima.ipac.caltech.edu/page/instruments}. For the Deep survey, the first 7 filters are always above the classical confusion limits, and it has been demonstrated that using those as priors the flux in the confused bands can be recovered within a $\sim20\%$ accuracy\citep{donnellan24}.   For \athena, we report the $2-10\rm{keV}$ $5\sigma$ sensitivities. }\vspace{1mm}
\begin{tabular}{cccccc}
\hline 
Survey & Area & Instrument   & t$_{\rm{field}}$(ks) & t$_{\rm{tot}}$ (hr)&Sensitivity  \\
\hline
\multirow{2}*{Deep} & \multirow{2}*{1 deg$^2$}& \textsl{PRIMAger} & & 1000  & $92-229\,\mu$Jy \\
& & \athena\ WFI & 300 & $>200$ & $1-2\times10^{-16}\,\rm{erg\,s^{-1}}$ \\
\hline
\multirow{2}*{Wide} &\multirow{2}*{28 deg$^2$} & \textsl{PRIMAger} & & 1000& $486-1211\,\mu$Jy \\
& & \athena\ WFI & 200 & $\sim4000$ & $1.5-3\times10^{-16}\,\rm{erg\,s^{-1}}$ \\
\hline
\end{tabular}
\label{tab:survey}
\end{table*}

\subsection{New Athena}\label{sec:ins_athena}

\athena\ is the upcoming ESA flagship X-ray observatory, designed to operate in the $0.2-12\,$keV energy range and address the "Hot and Energetic Universe" scientific theme \citep{nandra13}. The mission has recently completed a redefinition process to meet ESA's cost constraints, with the launch planned for the end of the 2030s. \athena\ features three key components in its scientific payload: an X-ray telescope with a 12-meter focal length, along with two instruments. These include the X-ray Integral Field Unit (X-IFU) \citep{barret20}, which will provide high-spectral-resolution imaging, and the Wide Field Imager (WFI) \citep{meidinger20}, designed for moderate-resolution spectroscopy over a large Field of View (FoV). 
The X-IFU on \athena\ will deliver simultaneous spatially resolved high-resolution X-ray spectroscopy, with a pixel size of 5 arcseconds and full-width at half maximum (FWHM) energy resolution of less than 4 eV below 7 keV, over a limited FoV of approximately 4 arcminutes in diameter. In contrast, the Wide Field Imager (WFI) will offer sensitive wide-field imaging and spectroscopy, with an energy resolution of FWHM $\leq 170$ eV at 7 keV, over a broader FoV and a wide energy range from 0.2 to 15 keV. The WFI achieves this using two sets of Silicon-based DEPFET Active Pixel Sensor detectors: the Large Detector Array, a 2$\times$2 mosaic covering a $\sim 40 \times 40,\text{arcmin}^2$ FoV, oversampling the point spread function (PSF) by more than a factor of two, and the Fast Detector, optimized for high count-rate observations.\par
In this paper, we focus on the \athena\ WFI, as it is specifically designed for large-area surveys. Its capabilities are expected to efficiently complement those of \prima, enabling a comprehensive study of obscured and Compton Thick (CT, those with $\nh\geq 10^{24}\,\rm{cm^{-2}}$)-AGN across multiple wavelengths.\par
The latest survey strategy envisioned for \athena\ is a ``wedding cake'' with 3 layers: a $12\,\rm{deg}^2$ area covered by $30\times300\,\rm{ks}$ pointings, a $28\,\rm{deg}^2$ area covered by $70\times200\,\rm{ks}$ pointings, and a $344\,\rm{deg}^2$ area covered by $860\times10\,\rm{ks}$ pointings (private communication). To properly study the synergies between \athena\ and \prima, we will compare the Deep $1\rm{\deg}^2$ \prima survey with $1\rm{\deg}^2$ of the \athena\ survey covered by $300\,\rm{ks}$ pointings (thus the maximum depth that \athena\ will deliver), and the $28\,\rm{deg}^2$ \prima\ Wide survey with the $70\times200\,\rm{ks}$ layer of the \athena\ survey (covering the same area). Table~\ref{tab:survey} summarises the survey strategy used in this work.

\section{Methods}\label{sec:methods}
\subsection{\athena\ and \prima\ detections} \label{sec:methods_detection}
To assess the capabilities of \athena\ and \prima\ in detecting AGN, we adopted the approach of \citet[][hereafter B21]{barchiesi21_spica}, modifying it to reflect the specifications of these new facilities. A detailed explanation of the method can be found in their work. \par
In summary, we began with the X-ray Background (XRB) synthesis model of \citet{gilli07} to estimate the total number of AGN as a function of redshift $z \in [0.3,10]$, intrinsic $2-10$keV luminosity $\Lx \in [42,48.2]$, and hydrogen column density $\Nh \in [20,26]$. By incorporating the observational capabilities of \athena, we determined the fraction of the XRB that could be resolved, while the unresolved fraction contributed to the background signal alongside the instrumental noise. A source was considered detected if it met a signal-to-noise ratio threshold of SNR $>5$, where the SNR is defined as the total number of counts divided by the background counts.\par
For each bin, we assigned 20 SEDs drawn from a sample of 422 AGN in the COSMOS field with X-ray counterparts, high AGN significance, and similar $\lx$ and $\nh$. The flux for each \textsl{PRIMAger} band was computed assuming a square transmission function. A source was considered to contribute to the detection fraction of its bin if its flux exceeded five times the sensitivity of the survey.\par
This process was repeated 42 times, selecting different SEDs in each iteration to ensure statistically robust estimates of the number of detected sources. The final detection estimates were taken as the median of these iterations, with uncertainties defined by the $16^{\rm{th}}$–$84^{\rm{th}}$ percentile range. Using 20 SEDs per bin, rather than one SED per expected detection, allowed us to optimize computational efficiency while minimizing the impact of stochastic SED selection, particularly in bins with a low number of expected sources.\par

In our simulations, we assumed that each detected source would also be recognised as an AGN. While characterising a source may be easy for the most luminous AGN or for those with X-ray coverage, it can be challenging for low-luminosity and very obscured AGN lacking \athena\ detections. We discussed this problem and how we can use \prima\ and a multi-wavelength approach to recognise AGN in Section~\ref{sec:discussions}.

\subsection{Black Hole Accretion Rate Density}\label{sec:methods_bhad}
To investigate the capabilities of \prima\ in constraining the BHAD evolution, we required the AGN bolometric luminosity function (LF) as it would be derived from \prima\ observations. We used the same $\nh$ binning as in Section~\ref{sec:methods_detection}; for the redshift, we chose 11 $z$ bins, using the same binning as in \citet[][hereafter D14]{delvecchio14} up to $z=3.8$ and extending it up to $z=10$. For the $\lx$ binning, we started from 21 $\lbolo$ bins in the range $9.5 \leq \Lbolosun \leq 19.5$ and converted those into $\lx$ bins via the Lusso et al. (2012, hereafter L12) bolometric correction\citep{lusso12}. We simulated a \prima\ deep observation by extracting $N$ SEDs in $\lx$, $\nh$, and $z$ bins, where $N$ is the expected number of objects (see Section~\ref{sec:methods_detection}).  For each filter, we removed the sources with flux lower than the survey sensitivity and used the remaining ones to calculate the LF via the non-parametric $1/V_{\rm{max}}$ method \citep{schmidt68}, where $V_{\rm{max}}$ denotes the maximum comoving volume within which each source is detectable. For each object, we determined the maximum redshift, $z_{\rm{lim}}$, at which detection is possible within the survey’s flux limits and computed the $V_{\rm{max}}$ as follows:
\begin{equation}
    \label{eq:vmax}
    V_{\rm{max}} = \int_{z_{\rm{min}}}^{z_{\rm{max}}} \frac{d V}{d z} \, \Omega(z) \,dz
\end{equation}
where $z_{\rm{min}}$ is the lower boundary of each $z$ bin and $z_{\rm{max}}$ is the minimum between the upper boundary and $z_{\rm{lim}}$. $\Omega(z)$ is the projected area of the survey corrected for completeness (computed as $1/f_c$, with $f_c$ the fraction of detected sources computed in section~\ref{sec:methods_detection}). \par
For each $\lbolo$ and $z$ bin, we computed the LF as follows:
\begin{equation}
    \label{eq:lf}
    \Phi(\lbolo,z) = \frac{1}{\Delta \log L_{bolo}} \sum_{i=1}^{n} \frac{1}{V_{\rm{max},i}} 
\end{equation}
with $\Delta \log L_{bolo}$ being the width of the $\lbolo$ bin. With a non-linear least square fitting algorithm, we fit the LFs with modified Schechter functions\citep{saunders90}:
\begin{equation}
    \label{eq:schechter}
    \Phi(L) \rm{d}\log L = \Phi^* \left( \frac{L}{L^*}\right)^{(1-\alpha)} \, \exp{ \left[ -\frac{1}{2\sigma^2} \, \log_{10}^2 \left( 1+ \frac{L}{L^*} \right) \right] } \rm{d}\log L
\end{equation}
where $L^*$ and $\Phi^*$ are the luminosity and the normalization of the knee of the LF, while $\sigma$ and $\alpha$ are the slopes of the LF below and above the knee. The modified Schechter function has 4 free parameters, as we have between 3 and 7 (depending on the $z$) bins to fit it, thus 0 to 3 degrees of freedom, we decided to fix the slopes to the best-fit values found by D14, i.e. $\alpha=1.48$ and $\sigma=0.54$.\par
Finally, the BHAD evolution is obtained from:
\begin{equation}
    \label{eq:bhad}
    \Psi_{\rm{BHAD}} (z) = \int_{0}^{\infty} \frac{1-\epsilon}{\epsilon \, c^2} \, \lbolo \, \Phi(\lbolo) \, d \log \lbolo
\end{equation}
with $\epsilon$ being the radiative efficiency of the SMBH, assumed to be $\epsilon=0.1$\citep{hopkins07,delvecchio18}.  While $\epsilon=0.1$ it is the standard common assumption \citep[e.g.,][]{hopkins07,delvecchio18,yang23}, we note that there is no general consensus on its value, with several works measuring $\epsilon$ as high as $\sim0.4$\citep{bian03,zhang20,farrah22}, while BH modelling predicts values between $0.06$ (for Schwartzschild BH) and $0.42$ (for a maximally rotating BH). For this work, assuming higher values would simply lower the normalisation of the BHAD, without changing its shape. Similarly to D14, we computed the uncertainties by performing the SED extraction and LF fitting 100 times. 

\section{Results}\label{sec:results}
In this section, we illustrate the results of our simulations. As an example, we show the capabilities of the first PRIMA Polarimeter Imager (PPI1) filter for both the Deep and Wide survey. We also present the number of expected photometric detections (i.e., the number of filters for which we expect to have $S/N>$ greater than the $5\sigma$ sensitivity of the survey) as function of $z$, $\nh$, and $\lx$ for both surveys. At the same time, we show which of these sources could be detected by \athena. As discussed in section~\ref{sec:methods}, our results illustrate the fraction of sources that can be detected by \prima\ and \athena, with the assumption that multi-wavelength coverage, SED-fitting, and spectroscopic follow-ups will correctly identify the sources as AGN and effectively constrain their bolometric luminosity. We refer to section~\ref{sec:discussions} for the discussion on the identification and characterisation of the sources. To highlight the advancements over previous instruments, we also compare our results with the detection capabilities of \textit{Herschel}. Specifically, we compare our predictions for the Deep survey with the values obtained by the Multi-tiered Extragalactic Surveys \citep[HerMES][]{olivier12} survey of the COSMOS field ($5\sigma$ sensitivity of $7.7\,$mJy at $100\,\mu$m). The Wide survey is compared to the XMM-LSS coverage from the same work ($5\sigma$ sensitivity of $49.9\,$mJy at $100\,\mu$m). These two surveys were selected as they represent the deepest available surveys over comparable areas. The overall numbers and fractions of expected AGN are reported in Tables~\ref{tab:deep_survey} and \ref{tab:wide_survey} for the Deep and Wide survey, respectively. \par
We also report the AGN bolometric LFs simulated from the Deep survey with filter PPI1, and their comparison with those measured by D14 using \textit{Herschel}-PACS data. Finally, we show the BHAD obtained by integrating the LFs and the comparison with those measured starting from X-ray or IR selections and with the values predicted by theoretical simulations.

\subsection{Deep survey}\label{sec:res_deep}
\begin{figure*}
  \centering
  \resizebox{0.75\hsize}{!}{\includegraphics{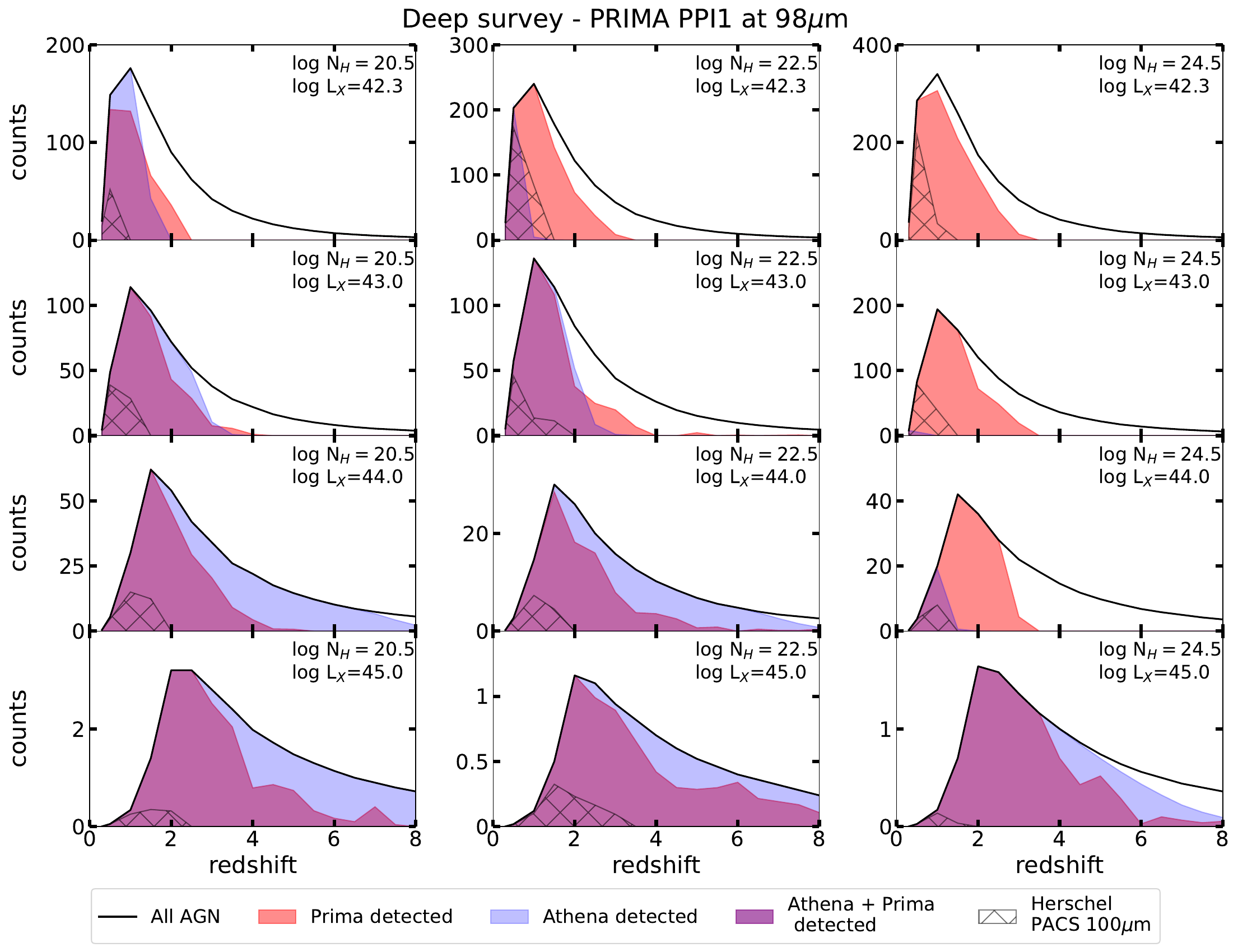}}
  \caption[Number of AGN expected per $\Delta z =1$ for the \prima\ Deep survey.]{Number of AGN expected per unit of redshift ($\Delta z =1$) for the \prima\ Deep survey. The black lines are the total number of expected AGN, the red areas represent those which can be detected with the \textsl{PRIMAger} PPI1 at 98$\mu$m, the blue areas those which can be detected in the X-rays with the \athena\ WFI, the purple areas are the AGN that will be detected by both instruments. For comparison, we reported (cross-hatched area) the performance of \textit{Herschel} PACS at $100\mu$m for a survey similar to the HerMES survey of the COSMOS field. The columns refer to AGN with different amount of obscuration (from left to right, $20 \le \lognh \le 21$; $22 < \lognh \le 23$; $24.18 < \lognh \le 25$), the rows to different AGN luminosity (from top to bottom, $42.0\le \loglx  <42.3 $, $42.9\le \loglx  <43.2 $, $43.9\le \loglx  <44.2 $, $44.9\le \loglx  <45.2 $).}
  \label{fig:deep_pp1} 
\end{figure*}

As an example of the PRIMA performance in the Deep survey, Fig.~\ref{fig:deep_pp1} illustrates the expected capabilities of \prima\ and \athena\ in detecting sources for 12 bins of $\lx$ and $\nh$. The black line represents the total number of AGN in each bin per unit of redshift. The red and blue areas are the fraction of sources that we will be able to detect with \textsl{PRIMAger} PPI1 (at $98\,\mu$m) and \athena\ WFI, respectively. The purple area is the fraction that should be detected with both instruments. Finally, the capabilities of the \textit{Herschel} survey of the COSMOS field are represented by the cross-hatched area.\par

As we can see, in this wavelength range, \prima\ will be able to detect almost all the AGN up to redshift 2. For $\loglx>43$, we also expect to have detections up to $z\sim4$. We can see that in this band, our ability to detect AGN depends mostly on the luminosity of the AGN, rather than their obscuration. On the other hand, for \athena\ we will detect most of the unobscured AGN up to $z\sim3$ (or $z\sim2$ for $\loglx \leq 42.3$) but miss most of the CT-AGN even at low-redshift. Where \athena\ shines is in detecting high-luminosity AGN, both obscured and unobscured: for $\loglx\geq44$ and $\lognh \leq 22.5$, we expect to detect all AGN up to $z\sim6-8$. Finally, we can see that \textit{Herschel} surveys are able to completely detect the AGN population only up to $z\sim 1$, and reveal the most luminous ones only up to $z \sim2$.\par
\begin{figure*}
  \centering
  \resizebox{0.75\hsize}{!}{\includegraphics{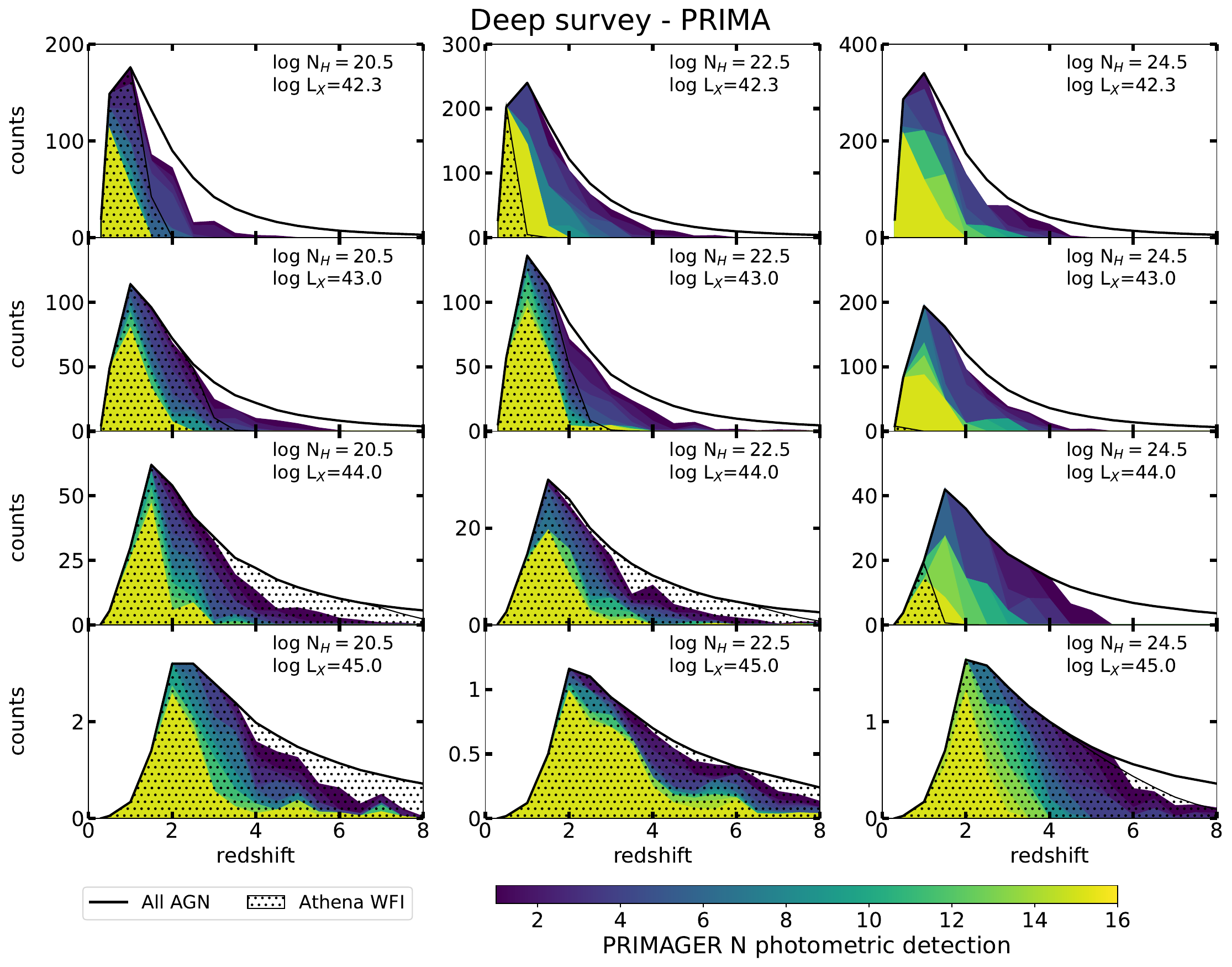}}
  \caption[Expected number of photometric detection per $\Delta z=1$ for the \prima\ Deep survey.]{Expected number of photometric detection per unit of redshift for the \prima\ Deep survey. The solid black lines represent the total number of expected AGN, while the dotted regions highlight those that will be detected in X-rays using the \athena\ WFI. The colour coding illustrates the number of \textsl{PRIMAger} filters capable of detecting these sources. Specifically, the yellow areas correspond to AGN detected across all filters, i.e. complete detections with both \prima\ PHI and PPI. In contrast, the dark blue regions represent sources expected to be detected in only one or two filters. Columns and rows follow the same scheme as in Fig.~\ref{fig:deep_pp1}.}
  \label{fig:deep_photometry} 
\end{figure*}

In Fig~\ref{fig:deep_photometry}, we illustrate the expected number of \textsl{PRIMAger} photometric detections for the Deep survey. The black lines are the total number of AGN per unit of redshift for different bins of $\lx$ and $\nh$. The dotted areas are the AGN fractions that we will be able to detect with \athena. The colour code indicates the number of \prima\ filters for which we expect to have a detection. For example, dark blue sources will be detected only in one or two photometric bands, while those in yellow will have complete coverage from $24$ to $235\mu$m. As we can see, we expect to be able to detect in all \textsl{PRIMAger} bands most of the $z\leq 2$ sources with $\loglx \geq 43$. For AGN with lower luminosity, a complete coverage is possible only up to $z\sim1$. However, we expect to be able to reveal all the AGN in at least a couple of filters up to $z\sim2$, effectively covering all the sources at the ``cosmic noon''. As most of the AGN lies at $z\leq3$, we expect \prima\ to detect $\sim30\%$ of all AGN in all the 16 bands and more than $70\%$ in at least one band. Furthermore,  we should be able to detect most AGN with $\loglx \geq 44$ in one or two photometric filters up to $z\sim 4$, with some detections expected up to $z\sim 6$. Finally, we stress that for the sources with a low number of filter detections but visible with \athena, the latter will allow us to recognise these sources as AGN and to put constraints on their properties (i.e., obscuration and intrinsic luminosities). On the other hand, multi-wavelength coverage, SED-fitting, and spectroscopic follow-ups can be exploited to fully characterise the sources without \athena\ detections (i.e., heavily obscured and low-luminosity AGN, see Section~\ref{sec:discussions}).

\par

A summary of the Deep survey performances of four of the \textsl{PRIMAger} filters, as well as of \athena\ and \textit{Herschel} is reported in Table~\ref{tab:deep_survey}.

\begin{table*}
\caption[Summary of the \prima\ and \athena\ capabilities in detecting AGN for the Deep survey.]{Summary of the \prima\ and \athena\ capabilities in detecting AGN for the Deep survey. $N_{\rm{AGN}}$ is the total number of expected AGN, WFI refers to the percentage that will be detected with \athena\ WFI. PHI1 and PHI2 refer to the percentage of AGN that we expect to detect with the two of the (sub-) filters of PHI (at $ 34\mu$m and $65\mu$m, respectively). Similarly for PPI1 and PPI3 (at $98\mu$m and $172\mu$m, respectively). For both PHI and PPI, the number in parenthesis is the fraction of sources that should be visible with both \athena\ and \prima. $N_{\text{filters}}$ is the average number of \textsl{PRIMAger} photometric filters that each source will be detected in (only for those detected). PACS refers to the fraction of sources that would be detected by \textit{Herschel} PACS at $100\mu$m for a sensitivity similar to the HerMES COSMOS survey.}\vspace{1mm}
\label{tab:deep_survey}
\centering
\small
\begin{tabular}{lcccccccc}
\hline
Deep survey & $N_{\rm{AGN}}$ & WFI & PHI1 & PHI2 & PPI1  & PPI3  & $N_{\text{filters}}$ & PACS \\
 & & \%& \%& \%& \%& \%&  & \% \\
All AGN & 25400 & 26 & $35\pm6\,  (17)$ & $34\pm4\,(16)$  & $59^{+4}_{-5}\,(23)$ & $70\pm4\,(25)$ & $7$ & 11\\
AGN $z \leq 4$ & 21600 & 29 & $41\pm7\,  (20)$ & $40\pm5\,(19)$ & $36\pm5\,(27)$ & $81^{+4}_{-5}\,(29)$ & $11$ & 14 \\
AGN $z \leq 2$ & 15200 & 33 & $52\pm8\, (25)$ & $54\pm6\,(25)$ & $85^{+4}_{-5}\,(33)$ & $92\pm3\,(33)$ & $14$ & 19\\
All CT-AGN & 10800 & 3 & $34\pm7\, (3)$ & $34\pm4\,(3)$  & $61\pm4\,(3)$ & $71\pm4\,(3)$ & $7$ &12 \\
CT-AGN $z \leq 4$ & 9200 & 4 & $40\pm8\,(4)$ & $40\pm4\,(4)$ & $71\pm5\,(4)$ & $83\pm4\,(4)$ &  $10$ &14 \\
CT-AGN $z \leq 2$ & 6600 & 5 & $49\pm9\,(5)$ & $55\pm6\,(5)$ & $87\pm4\,(5)$ & $93\pm2\,(5)$ & $14$ &20 \\
\hline
\end{tabular}
\end{table*} 

\subsection{Wide survey}\label{sec:res_wide}
\begin{figure*}
  \centering
  \resizebox{0.75\hsize}{!}{\includegraphics{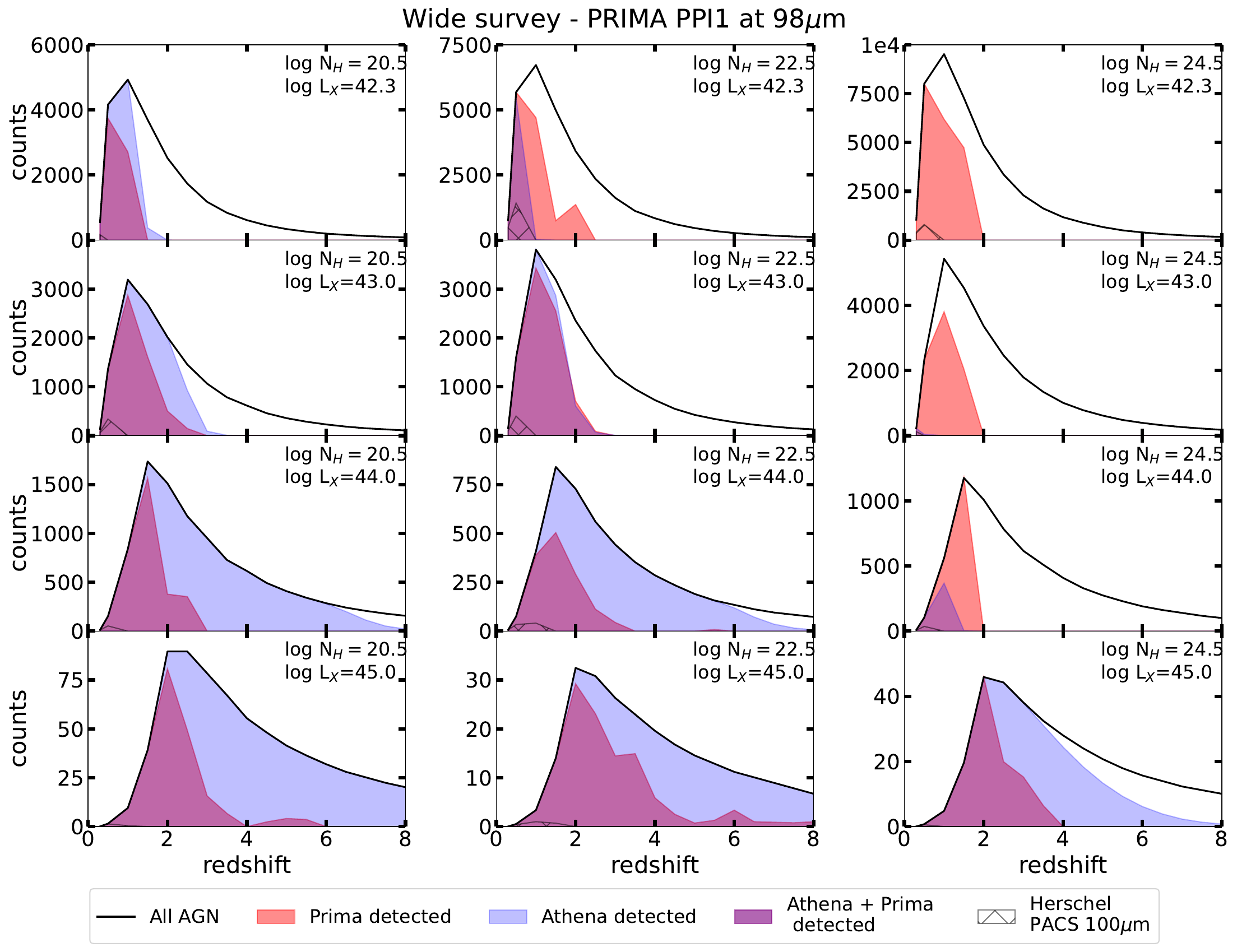}}
  \caption[Number of AGN expected per $\Delta z =1$ for the \prima\ Wide survey.]{Number of AGN expected per unit of redshift for the \prima\ Wide survey with PPI1. The lines and areas are coded as in Fig.~\ref{fig:deep_pp1}}
  \label{fig:wide_pp1} 
\end{figure*}

Similarly to sec~\ref{sec:res_deep}, we show the expected performance of \textsl{PRIMAger} PPI1 for the Wide survey in Fig.~\ref{fig:wide_pp1}. We used the same colour code as Fig.~\ref{fig:deep_pp1}, with the difference that the considered area (thus the AGN expected number) is 28 times larger.\par

As we can see, the capability of detecting high-$z$ AGN decreases significantly with respect to the Deep survey. In particular, we will be able to detect all the AGN only up to $z\sim1$, or $z\sim2 $ for those with $\loglx\geq 44$. On the other hand, the decrease in depth for \athena\ is not so impactful. This is due to the \athena\ sensitivity limit between the two surveys decreasing only by a factor $\sqrt{2/3}$. Similarly to the Deep survey, \athena\ on its own is not able to detect the majority of the CT-AGN (with the exception of the most luminous) and struggles with the obscured ($\nh \geq 10^{22.5}\,\rm{cm^{-2}}$) low-luminosity ones ($\lx \leq 10^{42.3}\,\rm{erg/s}$). Finally, we want to highlight that due to the large area (thus lower sensitivity), very few sources are detected by \textit{Herschel} PACS above $z\sim0.5$.\par
Despite the lower detection fractions, the Wide survey can take advantage of its $28\,\rm{deg^2}$ area to significantly boost the number of detected AGN. In particular, this survey can be extremely effective in revealing the rarest sources. For example, while we will not be able to detect all the AGN with $\loglx \sim 44$ and $\lognh\sim22.5$ (even those at $z\sim2$), we still expect to find $\sim14,000$ of them, a significant improvement over the $\sim1,200$ that the Deep survey will reveal. This is also evident in the number of expected detections of CT-AGN: while the Deep survey should reveal $\sim7,000$ of them, the Wide survey, although mostly limited to $z\leq2$, should boost this number by a factor $\sim10$.\par
\begin{figure*}
  \centering
  \resizebox{0.75\hsize}{!}{\includegraphics{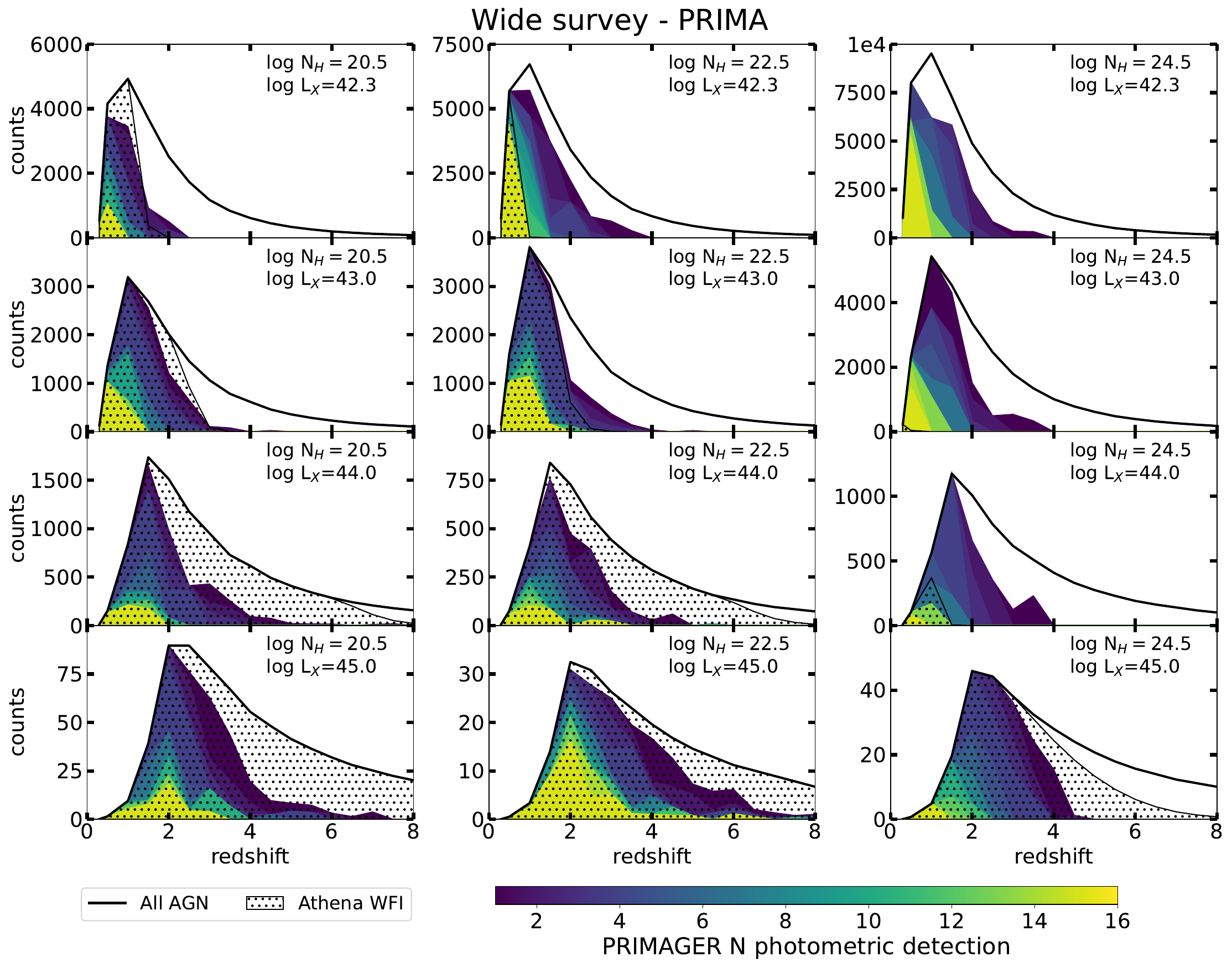}}
  \caption[Expected number of photometric detection per $\Delta z=1$ for the \prima\ Wide survey.]{Expected number of photometric detection per unit of redshift for the \prima\ Wide survey. The lines and areas are coded as in Fig.~\ref{fig:deep_photometry}.}
  \label{fig:wide_photometry} 
\end{figure*}

In Fig~\ref{fig:wide_photometry}, we illustrate the expected number of \textsl{PRIMAger} photometric detections for the Wide survey. We used the same colour code as in Fig.~\ref{fig:deep_photometry}. As we can see, we will be able to take advantage of all the bands only up to $z\sim1$, except for some of the most luminous AGN that we can be detected up to $z\sim2$. However, for the $z\leq 4 $ and $z\leq2$ sources that will be revealed (which should be more than $100,000$), we expect on average detections in 7 and 11 \textsl{PRIMAger} bands, respectively. Furthermore, as the differences between the depth of the Deep and Wide survey of \athena\ is minimal, we will be able to take advantage of the \athena\ detections to characterise the sources. In fact, with the exception of the most obscured sources and those with very low luminosities, most of the AGN with a couple of \prima\ photometric detections will also be revealed by \athena. We refer to Section~\ref{sec:discussions}, for a discussion on the characterisation of AGN lacking \athena\ detections.
We report in Table~\ref{tab:wide_survey} a summary of the \prima\ and \athena\ predictions for a Wide survey, as well as the comparison with \textit{Herschel} capabilities. 

\begin{table*}
\caption[Summary of the \prima\ and \athena\ capabilities in detecting AGN for the Wide survey.]{Summary of the \prima\ and \athena\ capabilities in detecting AGN for the Wide survey. Columns are the same as Table~\ref{tab:wide_survey}. PACS refers to the fraction of sources that would be detected by \textit{Herschel} PACS at $100\mu$m for a sensitivity similar to the HerMES XMM-LSS survey.}\vspace{1mm}
\label{tab:wide_survey}
\centering
\small
\begin{tabular}{lcccccccc}
\hline
Wide survey & $N_{\rm{AGN}}$ & WFI & PHI1 & PHI2 & PPI1 & PPI3  & $N_{\text{filters}}$ & PACS \\
 & $\times10^5$ & \%& \%& \%& \%& \%&  & \% \\
All AGN & $5.8$ & 24 & $11\pm3\,(6)$ & $15\pm3\,(8)$ & $36\pm4\,(15)$ & $47_{-6}^{+5}\,(18)$ & $4$ & 1 \\
AGN $z \leq 4$ & $5.0$ & 26 & $13\pm3\,(7)$ & $18_{-3}^{+4}\,(10)$ & $43_{-5}^{+4}\,(18)$ & $57_{-7}^{+6}\,(21)$ & $7$ & 2 \\
AGN $z \leq 2$ & $3.5$ & 30 & $19\pm5\,(10)$ & $25\pm5\,(13)$ & $59_{-6}^{+5}\,(25)$ & $75\pm7\,(27)$ & $11$ & 2 \\
All CT-AGN & $2.5$ & 2 & $13\pm3\,(2)$ & $16\pm3\,(2)$ & $37\pm3\,(2)$ & $47\pm5\,(2)$ & $6$ & 2\\
CT-AGN $z \leq 4$ & $2.1$ & 3 & $15\pm3\,(3)$ & $19\pm3\,(3)$ & $44\pm4\,(3)$ & $55\pm6\,(3)$ & $7$ &2 \\
CT-AGN $z \leq 2$ & $1.5$ & 3 & $21_{-5}^{+4}\,(3)$ & $27\pm5\,(3)$ & $61\pm5\,(3)$ & $74_{-8}^{+7}\,(3)$ & $10$ &3 \\
\hline
\end{tabular}
\end{table*} 

\subsection{Black Hole Accretion Rate Density Evolution}\label{ref:results_bhad}
\begin{figure*}
  \centering
  \resizebox{\hsize}{!}{\includegraphics{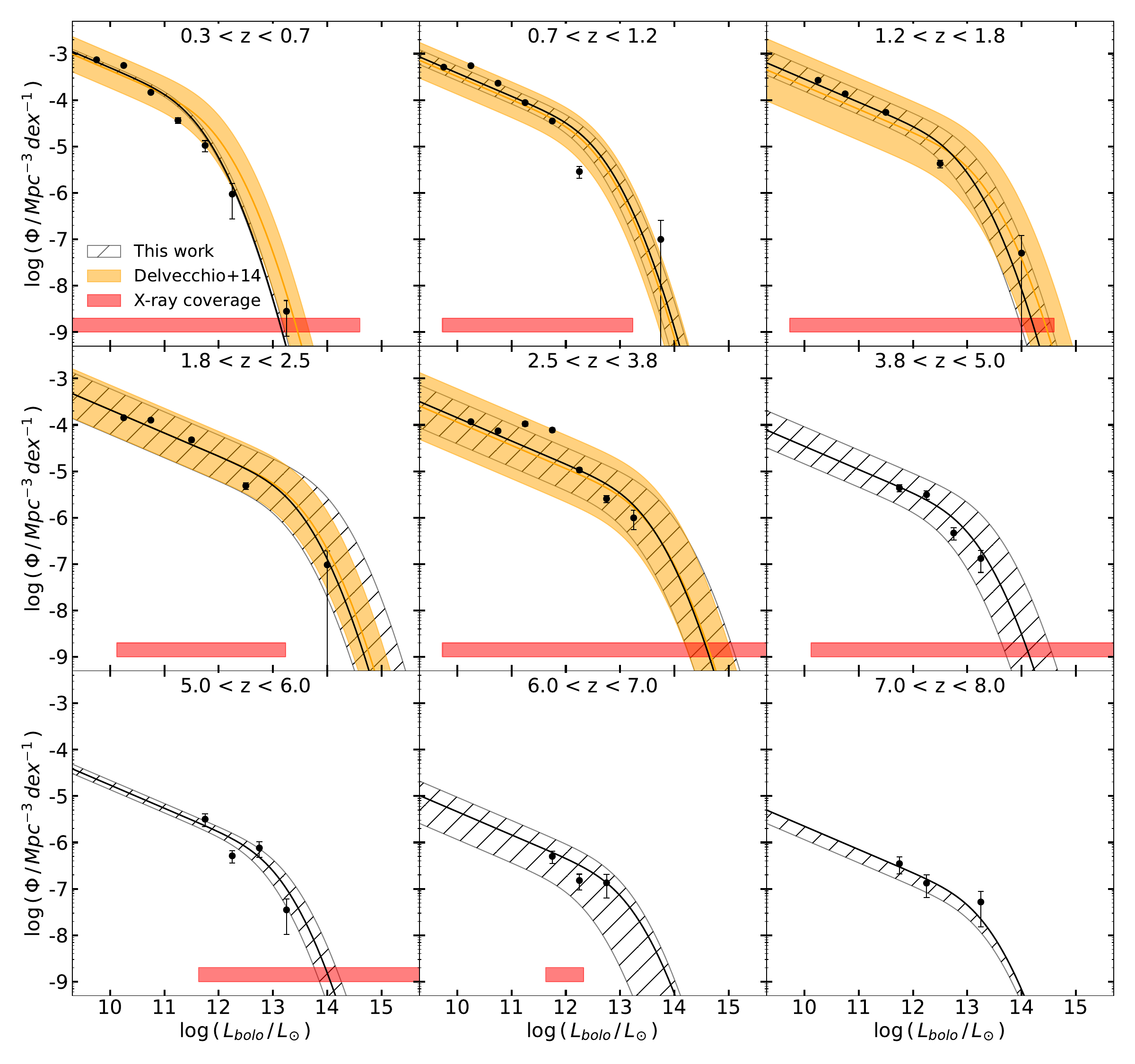}}
  \caption[AGN bolometric luminosity function.]{AGN bolometric luminosity function expected from a \prima\ Deep survey with PPI1 ($98\mu$m). The black line is our best-fit and the shaded area represents its uncertainties. The orange line is the best-fit LF derived from \textsl{Herschel}-PACS data by D14. The red areas highlight the bolometric luminosity range covered by current X-ray surveys\citep{aird15,vito18,pouliasis24}.}
  \label{fig:lf} 
\end{figure*}

Adopting the method described in section~\ref{sec:methods_bhad}, we were able to fit the LFs up to $z=8$. In Fig.~\ref{fig:lf}, we show the LFs obtained from PPI1 simulations in different $z$ bins. The error bars represent $1\sigma$ Poisson uncertainties \citep{marshall85}.The black line indicates our best-fit LFs, with the associated uncertainties (derived through random sampling of the SEDs; see Section~\ref{sec:methods_bhad}) shown as a hatched black area. For comparison, the orange line and shaded area denote the best-fit LF and its uncertainties derived by D14. The red-shaded region highlights the bolometric luminosity range accessible through current X-ray surveys\citep{aird15,vito18,pouliasis24}, assuming a L12 bolometric correction. Due to the low number of sources in the $7\leq z <8$ bin, we fixed the $L^*$ to the best-fit value found at $5\leq z <6$. We report in Table~\ref{tab:lf_params} the bolometric LF best-fit values and their $1\sigma$ uncertainties.\par

\par

\begin{table}
\caption[LF best-fit parameters]{List of best-fit parameters of the AGN bolometric LFs and associated uncertainties. We fitted the LFs with modified Schechter functions, fixing the faint- and bright-end slopes to the values of D14, i.e., $\alpha=1.48$ and $\sigma=0.54$. For the $7\geq z < 8 $ bin, due to the low number degrees of freedom, we also fixed the luminosity at the knee of the LF at the values of the previous bin.}\vspace{1mm}
\label{tab:lf_params}
\centering
\small
\begin{tabular}{ccc}
\hline
$z$ bin & $\log{(L^*\,/\,\rm{L_{\odot}})}$ & $\log{(\Phi^*\,/\,\rm{Mpc^{-3}dex^{-1}})}$  \\
$0.3 \leq z < 0.7$ & $11.00_{-0.01}^{+0.04}$ & $-3.8_{-0.05}^{+0.03}$ \\
$0.7 \leq z < 1.2$ & $12.0_{-0.2}^{+0.1}$ & $-4.36_{-0.08}^{+0.09}$ \\
$1.2 \leq z < 1.8$ & $12.2_{-0.2}^{+0.3}$ & $-4.87_{-0.18}^{+0.12}$ \\
$1.8 \leq z < 2.5$ & $12.2_{-0.1}^{+0.6}$ & $-4.9_{-0.5}^{+0.1}$ \\
$2.5 \leq z < 3.8$ & $12.9_{-0.2}^{+0.4}$ & $-5.3_{-0.5}^{+0.1}$ \\
$3.8 \leq z <  5$ & $12.6\pm0.4$ & $-5.9\pm0.2$ \\
$5 \leq z < 6   $ & $12.2\pm0.2$ & $-6.0\pm0.1$ \\
$6 \leq z < 7   $ & $12.7_{-0.7}^{+0.2}$ & $-6.5\pm0.3$ \\
$7 \leq z < 8   $ & $12.7$ & $-0.7_{-0.3}^{+0.1}$ \\
\hline
\end{tabular}
\end{table}

Our LF results are consistent with those of D14 across all redshift bins. The most significant differences are seen in the lowest redshift bin, where we have fewer sources at the bright-end of the LF and small uncertainties. These small uncertainties are mainly driven by two factors:
firstly, \prima\ detects almost all the sources in the entire redshift bin, thus there is little difference in extracting one SED or another; secondly, we expect $\sim1000$ sources but have only 400 SEDs at our disposal, thus the variations when bootstrapping are minimal.\par
We further examined the points showing the largest deviation from our LF best-fit (e.g., $\Lbolosun \sim12$ at $0.7 \leq z < 1.2$ and $\Lbolosun \sim 11-12$ at $2.5 \leq z < 3.8$). Their behaviour can be attributed to the binning used in computing the LF. In particular, it arises from simulating the expected number of sources and the survey completeness (see section~\ref{sec:methods}) in quite large $\lx$ and $z$ bins that do not exactly match the $\lbolo$ and $z$ binning used to compute the LF.\par

\begin{figure}
  \centering
  \resizebox{\hsize}{!}{\includegraphics{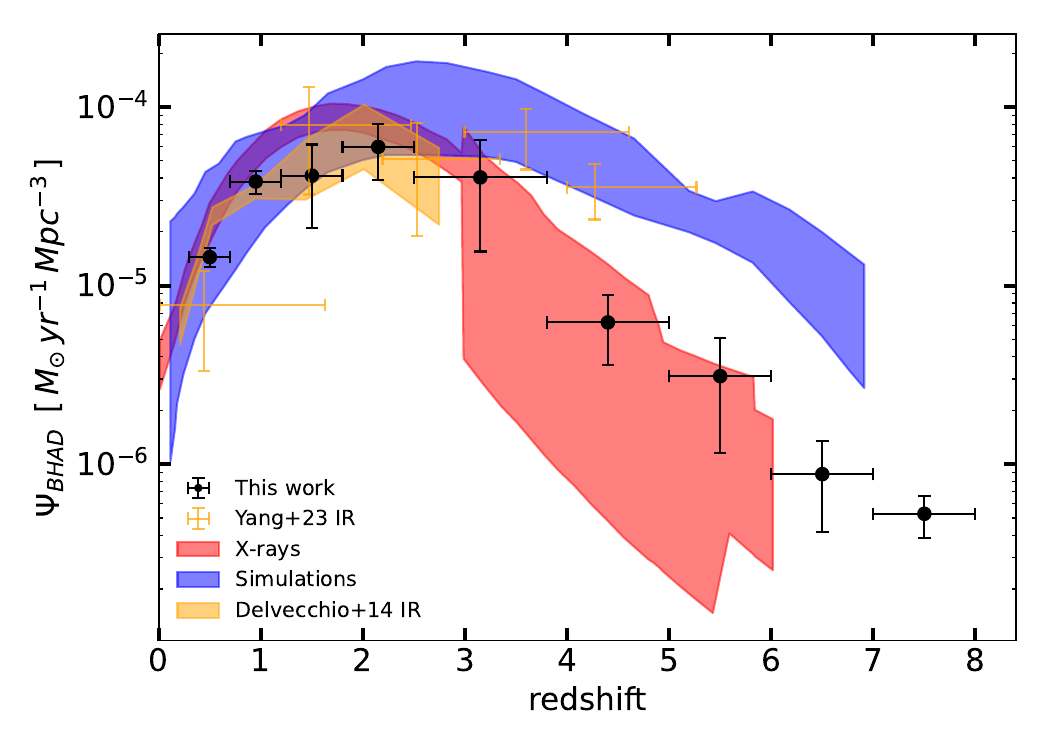}}
  \caption[Prediction of \prima-derived measurements of the BHAD.]{Prediction of \prima-derived measurements of the BHAD. The black points represent the predictions of the BHAD obtainable with a \prima\ Deep survey with PPI1 ($98\,\mu$m). For comparison, we report the BHAD measured from X-ray surveys \citep[red area][]{ueda14,vito14,aird15,vito18,pouliasis24}, and IR surveys \citep[orange area and points][]{delvecchio14,yang23}. In blue, BHAD predictions from simulations \citep{volonteri16,sijacki15,shankar13} are shown. Our points follow the BHAD measured from the X-rays as our simulations started from X-ray background modelling. }
  \label{fig:bhad} 
\end{figure}

In Fig.~\ref{fig:bhad}, we demonstrate the capabilities of \prima\ in measuring the BHAD. The black points are our simulated measures of the BHAD obtained by integrating the LFs (see section~\ref{sec:methods_bhad}). For comparison, the red-shaded area represents measured BHAD values from X-ray-selected galaxies \citep{ueda14,vito14,aird15,vito18,pouliasis24}. Predictions from various simulations are shown as a blue-shaded region \citep{shankar13,sijacki15,volonteri16}, while the orange-shaded area and points correspond to BHADs measured from \textsl{Herschel}-PACS selected sources in D14 and JWST-selected sources \citep{yang23}, respectively. Up to $z\sim3$, our predictions align with the BHAD values from D14 and are consistent with those measured via X-ray observations. At $z>3$, our BHAD estimates follow the trends observed in X-ray surveys, which is expected as our simulations started from X-ray background modelling. The smaller uncertainties at $7\leq z <8$ (with respect to the previous bin) arise from fixing the $L^*$ in fitting the LF.  \par
Regardless of the simulated BHAD values at high redshifts, we find that with \prima\ we will be able to measure the evolution of the BHAD up to $z\sim8$ with reasonable accuracy. As a sanity check, we utilized the SPRITZ simulations of \citet[][hereafter B24]{bisigello24} to evaluate the capabilities of \prima\ in measuring the BHAD. Briefly, SPRITZ derives the number densities of star-forming galaxies, AGN, and composite objects up to $z=10$ from the IR LFs of these populations. The simulated IR luminosity range span between $L_{IR}=10^{5}\,\rm{L_{\odot}}$ and $L_{IR}=10^{15}\,\rm{L_{\odot}}$. Each object in the simulation is assigned a SED model based on its classification, which is then used to derive fluxes in various bands. For further details, we refer to B24. We used the same approach of sec.~\ref{sec:methods_bhad} to measure the bolometric LFs and BHAD derived from the $100\mu \rm{m}$ detections. We found that \prima\ is able to recover more than $90\%$ of the simulated BHAD at $z\geq1$ and up to $z\sim8$. The BHAD simulated with SPRITZ is systematically higher than that obtained from our simulations and remains nearly constant at $z>3$, similarly to  JWST-measured BHAD \citep{yang23}. Due to the intrinsically higher BHAD predicted by SPRITZ, its measurements exhibit even smaller uncertainties than those from our simulations: approximately $\sim5\%$ at $z\leq3$, $\sim 10\%$ at $4\leq z\leq6$, and $\sim35\%$ at $z \geq 7$. It is important to emphasize that, regardless of the actual evolution of the BHAD, the objective of this work is to demonstrate that \prima\ will be capable of measuring it with reasonable accuracy. Given that the source number densities in B24 are derived from IR LFs rather than from the XRB as in our work, and that different SED models were employed, our findings provide strong evidence that \prima\ will effectively be able to measure the BHAD at high redshifts. 

\section{Discussion}\label{sec:discussions}
Our simulations of the capabilities of \prima\ and \athena\ are reported in section~\ref{sec:results}, and summarised in Tables~\ref{tab:deep_survey} and \ref{tab:wide_survey}. Figure~\ref{fig:bhad} shows \prima\ capabilities in reconstructing the evolution of the BHAD. In this section, we discuss the synergies between the two instruments, the improvements over existing surveys, and the huge importance that FIR observatories such as \prima\ will have in the next decades.\par

As shown in Fig.~\ref{fig:deep_pp1} and Fig.~\ref{fig:wide_pp1}, the two layers of the \prima\ survey strategy ($1000\,\rm{hr}$ covering $1\,\rm{deg^2}$ and $1000\,\rm{hr}$ covering $27\,\rm{deg^2}$) complement each other effectively. The Deep layer will detect nearly all AGN up to $z \sim 2$ (at least down to $\Lbolo\sim43$, i.e., the lower limit in our simulations), and will detect sources up to $z \sim 4$ and beyond (those with $\Lbolo \geq 45$), providing excellent (and, for sources with $\Lbolo \geq 45.5$, nearly complete) coverage of the ``cosmic noon''. Meanwhile, the Wide survey, with its large area, will excel in the number of detectable sources. We expect to detect up to approximately $3 \times 10^5$ AGN in total with $\Lbolo\sim43.3$, of which about 30,000 with complete detections across all 16 \textsl{PRIMAger} bands. Multiple detections across various bands are critical for two reasons. First, shorter wavelength detections can help resolve sources that are blended at longer wavelengths, a technique successfully demonstrated using both \textit{Spitzer} and \textit{Herschel} observations \citep{hurley17,wang24}, as well as \prima\ simulations \citep{donnellan24}. Secondly, in the absence of spectroscopic data, the most effective way to characterise a source and derive its properties is through SED fitting, and \prima\ will excel at it\citep{bisigello24}. Given the degeneracies between AGN and host-galaxy models and their intrinsic uncertainties, the ability to separate the two components is directly linked to the number of photometric detections. Currently, the best FIR coverage comes from \textit{Spitzer} and \textit{Herschel}, but for most sources at $z \sim 1$ and beyond, there are typically only one or two photometric detections. \prima, with its 16 contiguous filters, will be a game changer in SED-fitting of high-$z$ objects. Fig.11 of B21 demonstrates the improvement in constraining the AGN and host galaxy properties provided when incorporating four additional FIR filters. B24 showed that we can effectively recognise AGN by using the \prima\ coverage to measure the shape of the IR dust continuum. B24 simulated \prima\ observations of ``normal'' and active galaxies over a large range of AGN fractions, IR luminosities, PAH contribution, dust properties, and redshifts and performed SED-fitting to investigate \prima\ capabilities in recovering the source properties. In particular, exploiting PHI, PPI1, and PPI2 filters, B24 were able to retrieve the AGN fraction with a dispersion of 0.06, the fraction of dust mass in PAH with a precision of $\sim10\%$, and the IR luminosity with a scatter of 0.1 dex. \par
Additionally, most of the obscured AGN emission is usually in the $5-30\,\mu$m (rest-frame) band coinciding with the wavelength range where PAH and dust emission from the host galaxy are also present. Properly disentangling these components requires both good coverage at these wavelengths and longer-wavelength photometric detections to constrain the host-galaxy dust emission, both of which will be provided by \prima. While current facilities like JWST cover up to $28\,\mu$m (albeit JWST is not ideal for large-area surveys), much of the obscured AGN emission is shifted out of its bands for $z \geq 1$. In contrast, \prima\ continuous coverage from 24 to $250\, \mu$m can trace the AGN emission up to $z\sim10$ and beyond, and its ability to detect high-$z$ AGN will mainly depend on the survey depth.\par
\begin{figure*}
  \centering
  \resizebox{\hsize}{!}{\includegraphics{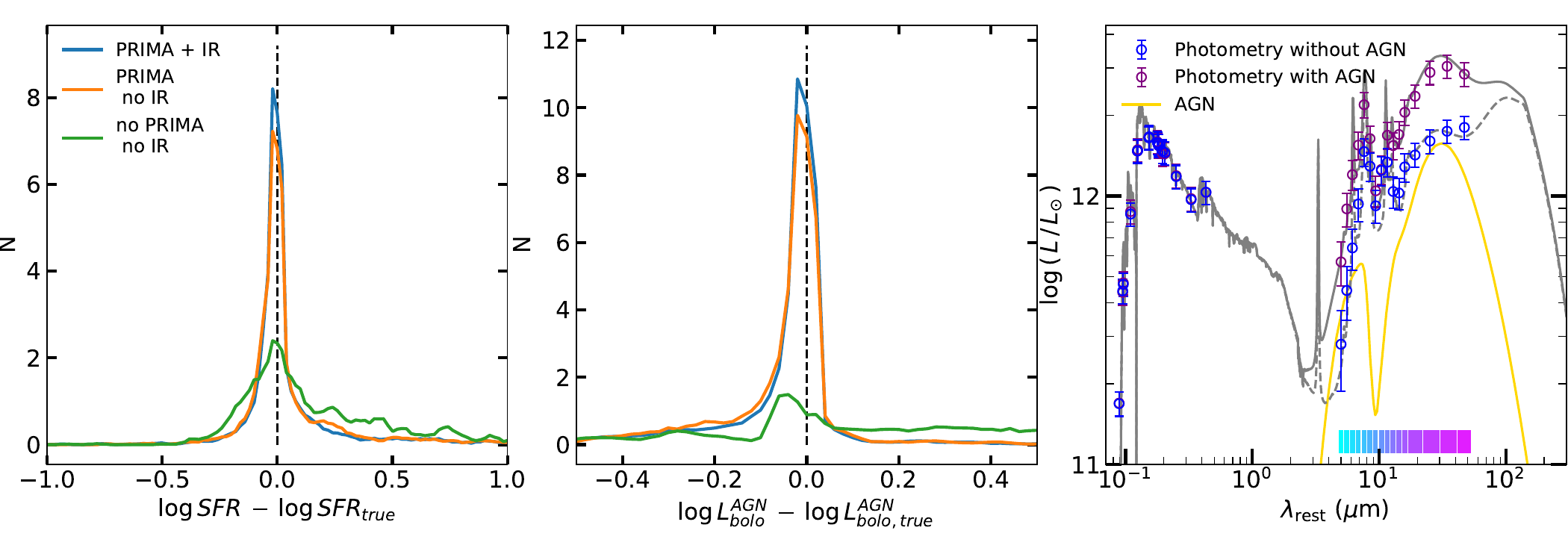}}
  \caption[Distribution of the difference between the estimated and the true values of the extracted parameters.]{\textit{Left} and \textit{center panels}: Normalised distribution of the difference between the estimated and the true values of the SFR and AGN bolometric luminosity. The blue lines show the result from the SED-fitting using all the available filters; the orange lines represent those using only the \prima\ PHI and PPI bands. Finally, the green lines show the results for sources without any mid- and far-IR detection. \textit{Right panel}: Example SEDs of a galaxy with and without an AGN. The grey dashed line represents the total SED of a non-active galaxy at $z\sim4$  with $\log{(M_*/\rm{M_{\odot}})}=10.1$ and $SFR=150\,\rm{M_{\odot}\,yr^{-1}}$. The solid grey line corresponds to a similar galaxy hosting an AGN with $\Lbolo = 45.6$, whose contribution is shown by the yellow line. Simulated photometric points, including \textsl{CFHT, MegaCAM, SUBARU, Vista, Spitzer/IRAC}, and \prima\ filters, are displayed in blue (without AGN) and purple (with AGN). The colour bar at the bottom indicates the wavelength coverage of \prima\ PHI and PPI bands. Despite the AGN having lower flux compared to the host galaxy dust emission in most \prima\ bands, the extensive coverage of \prima\ enables to effectively separate the two contributions, allowing for an accurate measurement of the AGN bolometric luminosity.}
  \label{fig:sedfitting_prop} 
\end{figure*}

 To accurately reconstruct the BHAD, it is essential not only to identify sources as AGN, but also to precisely measure their bolometric luminosities. Therefore, we further assessed our ability to recover AGN and host-galaxy properties through SED-fitting. Following the methodology of B24, we used CIGALE to simulate observations of high-redshift AGN. We considered six redshift bins with $1\leq z\leq 6$, generating 1280 SEDs per redshift bin. These SEDs encompass a range of star formation histories, stellar ages, AGN fractions, and AGN inclinations. Specifically, we simulated sources with stellar masses of $M_* = 5 \times 10^{10}$ and $5 \times 10^{11}\,\rm{M_{\odot}}$, star formation rates $\log{(SFR\,/\,\rm{M_{\odot}\,yr^{-1}})\in[-5,4]}$, and bolometric luminosities $\Lbolo \in [43,48]$. We then performed SED fitting on the simulated observations and compared the recovered properties to the intrinsic values used to generate them. The simulated observations were designed to resemble real deep-field surveys. The optical-NIR coverage is identical to that of the COSMOS 2020 catalog \citep{cosmos2020}, while the mid- and far-IR data are modeled after the XID+ deblended catalog \citep{wang24}. For each band, we adopted a sensitivity corresponding to the median uncertainties of the sources in these catalogs. Additionally, we simulated \prima\ PHI and PPI observations, incorporating uncertainties equivalent to the $1\sigma$ survey sensitivity. Gaussian noise, with a standard deviation matching the band uncertainty, was added to all photometric fluxes. The left and center panels of fig.~\ref{fig:sedfitting_prop} present the normalized distribution of the difference between the SED-fitting derived and true values for the $SFR$ and the $\lbolo$. We examine three scenarios: (i) without mid- and far-IR coverage (a common situation for $z>3$), (ii) including \prima\ coverage, and (iii) incorporating \prima, along with \textsl{Spitzer} and \textsl{Herschel} data. Notably, complete wavelength coverage does not necessarily imply detections in all bands; if the flux plus noise fell below the band sensitivity threshold, the corresponding photometric point was treated as a non-detection. We found that with \prima\ coverage we can recover (within 0.1 dex of the true value) the bolometric luminosity for $75\%$ of our sources and for $63\%$ of those at $z\geq3$, a significant improvement over the $18-20\%$ of the sources without mid-IR and far-IR coverage. We also expect \prima\ to double the number of sources with reliable SFR (within 0.1 dex of the true value), with the fraction increasing from $\sim32\%$ to $63\%$. We run further simulations focusing on sources at $6\leq z \leq 8$. We found that using \prima\ photometric bands, we measure bolometric luminosity within 0.1 dex of the true value for $\sim30\%$ of the simulated objects. More importantly, for the majority of sources with $\Lbolo\geq 45$, the measured $\lbolo$ is within 0.3 dex of the true value. Specifically, for sources with $45\leq \Lbolo <46$, we obtain an average offset of $\log{L_{\rm{bolo}}^{\rm{sed}}}-\log{L_{\rm{bolo}}^{\rm{true}}} = 0.24 \pm 0.30$, similar to the bayesian uncertainties associated to the measured values. Considering the last two redshift bins in Fig.~\ref{fig:lf} ($6\leq z <7$ and $7\leq z <8$), we can see that all the bolometric luminosity bins that we are able to populate with \prima\ are at $\Lbolosun\geq 11.5$, or equivalently to $\Lbolo \geq 45$. Therefore, we are confident that \prima will not only be able to detect sources up to $z\sim8$, but also to measure their $\lbolo$ with reasonable accuracy, enabling us to reconstruct the bolometric LFs and BHAD even at these high redshifts. \par

Regarding \athena\, as its first two layers have similar depths (with $t_{\rm{field}}=300$ and $200\,$ks, respectively), the fractions of sources detected by the Deep and Wide surveys differ only by some percent. In general, we expect \athena\ to detect $\sim25\%$ of all the AGN, and over $30\%$ of those at $z\leq2$. In particular, \athena\ will be extremely powerful in detecting moderate-luminosity AGN up to $z\sim6$ and the high-luminosity ones up $z=8$ and beyond. For unobscured low-luminosity AGN, \athena\ should be able to detect all of them up to $z\sim1$ and most of them at $z\sim2$. However, from Fig.~\ref{fig:deep_photometry}, it is easy to see that the capabilities of \athena\ are heavily affected by the source intrinsic obscuration. While it can reveal AGN with $\lognh \sim 20.5$ and $\loglx\sim42.3$ up to $z\sim2$, if we increase the obscuration to $\lognh=22.5$ we are not able to go beyond $z\sim1$. The difficulty in detecting heavily obscured AGN is easily noticeable for CT-AGN: at high-$z$ \athena\ can reveal only the most luminous, those with $\loglx \leq 44$ are not visible above $z\sim1.5$, and AGN with moderate and low luminosity ($\loglx \leq 43$) will be completely missed. As low- and moderate-luminosity AGN comprises most of the CT-AGN population, the fraction of \athena\ detections of CT-AGN is only $2-3\%$.\par
The synergies between \athena\ and \prima\ are particularly significant in the study of obscured and CT-AGN. Many CT-AGN that will remain undetected by \athena\ will likely be observed by \prima\ (at least up to $z \sim 3$). These CT-AGN are expected to play a crucial role in the evolution of the BHAD and may help resolve the gap between the measured BHAD and theoretical predictions from simulations. Currently, most BHAD estimates are derived from X-ray surveys, and, as such, suffer from obscuration. Even with the advanced capabilities of next-generation X-ray observatories like \athena, a large portion of CT-AGN will remain undetected without complementary surveys in the FIR, such as those provided by \prima. The excellent capability of \prima\ to recover the BHAD at very high redshifts is clearly demonstrated in Fig.~\ref{fig:bhad}.  Conversely, X-ray coverage of FIR-detected sources will be essential for characterising these objects and distinguishing between obscured AGN and dusty SF galaxies. \athena\ will provide both imaging and spectroscopy, enabling constraints on key X-ray properties, including $\lx$ (and consequently AGN bolometric luminosities) and $\nh$. In addition, the intrinsic X-ray luminosity (i.e., corrected for the obscuration) can be fed to SED-fitting codes to enhance our ability to accurately fit the optical and IR regimes and separate the AGN and host-galaxy contributions. Since \athena\ and \prima\ should fly within the same timeframe, it is not unreasonable to expect a coordinated survey strategy that will fully exploit the synergies between these two instruments.\par
The improvement of \prima\ over existing surveys is easily seen in Fig.~\ref{fig:deep_photometry} and even more in Fig.~\ref{fig:wide_photometry}. From our simulations, the \textit{Herschel} survey of the COSMOS field reached sensitivity that allowed us to detect $\sim 10\% $ of all the AGN and $\sim 20\% $ of those at $z\leq2$. 
At the same band ($\sim 100\mu$m), the \prima\ Deep survey is expected to reveal $\sim60\%$ of all AGN and $\sim85\%$ of those at $z\leq 2$. In addition, $\sim30\%$ of all AGN should have complete detection in all 16 \textsl{PRIMAger} bands. \prima will not only detect all of the AGN at the redshifts that \textit{Herschel} was able to reach, but it will also push its detections well beyond the ``cosmic noon''. Therefore the number of detected AGN is expected to improve by a factor $3-8$. The improvement is even stronger when considering the Wide survey. For such a survey, \textit{Herschel} was not able to detect more than $1-2\%$ of the AGN populations, while \prima\ will reveal between $\sim10$ and $\sim50\%$. Summarising, regarding galaxy evolution and AGN, \prima\ can be considered a true successor of \textit{Herschel}: it will sample a similar parameter space, but with better sensitivity and more photometric bands, thus allowing us to characterise the entire population of the sources detected by \textit{Herschel} and to reach redshifts that were not possible before. \par

Regarding the BHAD, our simulations demonstrate that \prima\ will be able to measure it up to $z\sim8$, regardless of the BHAD assumed in this work and of the SED utilised. Prior to the advent of JWST, the deepest IR-derived BHAD measurements extended only to $z\sim3$. Currently, we are able to measure it up to $z\sim4-5$, though JWST is not ideal for conducting large-area surveys. With current X-ray surveys, we can measure the BHAD up to $z\sim6$, however, a major challenge is correcting for the contribution of the most obscured AGN, that are mostly missed by X-ray selections. As already discussed, \prima\ detection capabilities do not depend on the source obscuration and its completeness is mostly dominated by the source fluxes, making it far easier to compute. Our simulations show that \prima\ capabilities in measuring the BHAD will exceed any current IR or X-ray instrument. With \prima, we will be able to explore the BHAD at previously inaccessible redshifts and address long-standing questions, such as whether the gap between X-ray-derived BHAD estimates and theoretical predictions is caused by the contribution of highly obscured AGN.\par
Accurately measuring the BHAD across cosmic times not only enables the study of black hole accretion mechanisms and feedback processes but could also provide a potential test for the dark energy content of the Universe. As the local SMBH function must be equal to the integral of the AGN accretion across cosmic times plus the total mass increase due to cosmological coupling (i.e., the BH mass growth due to the cosmological expansion), knowing the BHAD evolution and the local SMBH would allow us to measure the cosmological coupling. The latter could provide insight on the equation of state of the matter inside the BH and on their possible contribution to the dark energy content of the Universe\citep{soltan82,cadoni23,lacy24}.\par
While we compared our BHAD predictions with those derived from X-ray and infrared surveys, as well as with estimates from cosmological simulations, recently there have been attempts to measure it starting from JWST detections of dust-reddened broad-lines AGN, commonly referred to as ``Little Red Dots'' \citep[LRDs][]{matthee24}. While the AGN nature LRDs is still a matter of discussion, with alternatives comprising compact dust-rich star-forming galaxies, brown dwarfs, or other more exotic sources \citep{labbe23,perez-gonzalez24,willians24,akins24}, assuming an AGN origin lead to $z\sim6-8$ BHAD measurements that exceed any of the previous of at least one order of magnitude \citep{inayoshi24}. This additional discrepancy further strengthens the critical need for accurately measuring the BHAD at high-$z$. Our predictions of the BHAD are inherently dependent on the assumptions we made in our simulations. In particular, on the total number of expected AGN and its redshift evolution, which we based on the XRB modelling. As discussed in section~\ref{ref:results_bhad}, assuming a different BHAD does not impact significantly our capability of recovering it via \prima\ surveys. While the exact values of the BHAD depend on the assumed distribution of AGN and their evolution, the uncertainties associated to our simulation do not. Instead, they reflect the intrinsic BHAD uncertainties measurable with \prima. The legacy value of this work is in showing that \prima\ will allow us to measure the BHAD with better accuracy than current IR and X-ray surveys, up to higher redshift than what is now possible, and free from obscuration biases.\par
To facilitate direct comparisons with PACS-derived BHAD measurements, this work has focused on demonstrating \prima\ capabilities in measuring the BHAD specifically starting from a $100\,\mu$m survey. We want to highlight that \textit{PRIMAger} will observe simultaneously across its entire wavelength range, from $24\,\mu$m to $261\,\mu$m, offering several advantages. First, lower-frequency detections with \textit{PRIMAger} can be used as priors to deblend sources at longer wavelengths, ensuring more reliable flux measurements \citep{donnellan24}. Second, as discussed previously, it will allow us to characterise the host galaxy and AGN properties, facilitating robust AGN identification \citep{barchiesi21_spica,bisigello24}. Lastly, having contemporary surveys at all \prima\ wavelengths will allow us to merge those to increase the number of detected AGN. This is particularly critical for identifying obscured AGN in quiescent or faint-FIR galaxies that may remain undetected at $100\,\mu$m. The increased AGN sample will simplify the incompleteness corrections, further reducing the BHAD uncertainties.\par 
Characterising newly discovered sources will strongly depend on accurately constraining their redshifts. While reliable photometric redshifts should be achievable for AGN detected across all \textsl{PRIMAger} bands, sources with a low number of photometric detections will require spectroscopic follow-ups. However, as in the next decades we expect a huge number of new objects to be revealed by telescopes such as LSST, SKAO, CTA, and the Nancy Grace Roman Space Telescope, new facilities and surveys dedicated to efficiently provide spectroscopy are also being developed. In particular, the \prima\ surveys could be conceived to cover fields that will have complete deep spectroscopic coverage at the time of its launch. For example, in the next years, the COSMOS, GOODS, and Euclid Deep Fields will receive spectroscopic coverage from 4MOST. The Wide-Area VISTA Extragalactic Survey (WAVES) and the Optical, Radio Continuum, and HI Deep Spectroscopic Survey (ORCHIDSS) will target more than 180,000 galaxies in these fields.  Additionally, \prima\ could benefit from planned facilities like the Wide Field Spectroscopic telescope (WFS), which is expected to deliver $\textrm{R}=3000-4000$ Multi-Object Spectroscopy for 250 million sources, along with over 4 billion spectra via its $\textrm{R}=3500$ Integral Field Spectrograph \citep{bacon24}.\par
Finally, we highlight that, while its investigation exceeds the scope of this work, \prima\ will also carry on board a pointed spectrometer (\textsl{FIRESS}) covering the $24-235\mu$m wavelength range and able to operate in low- ($\textrm{R}\sim100$) and high-resolution ($\textrm{R}\sim 4400-12000$) modes. In the case of sources with few photometric detections and not visible by \athena\ (e.g., most of CT-AGN), characterising these sources and constraining their host-galaxy and AGN properties will benefit from follow-ups with \textsl{FIRESS}. In particular, AGN-related high-ionisation lines, such as [Ne$\,\textsc{v}$]$14.3\mu$m and $24.3\mu$m, [O$\,\textsc{iv}$]$25.9\mu$m \citep{tommasin08, tommasin10, feltre16}, and the $9.7\mu$m and $18\mu$m silicate features will fall within the \textsl{FIRESS} wavelength coverage. The $9.7\mu$m feature is typically associated with unobscured AGN when observed in emission and with obscured ones when it is in absorption. Moreover, its depth has been shown to correlate with the X-ray derived $\text{N}_{\text{H}}$\citep{wu09,shi06,Goulding12}. To quickly illustrate the capability of FIRESS in performing follow-up observations of AGN detected by \textsl{PRIMAger}, we utilised $L_{\rm{bol}}-L_{\rm{line}}$ relations to estimate the fluxes of the [Ne$\,\textsc{v}$]$14.3\mu$m and $24.3\mu$m, [O$\,\textsc{iv}$]$25.9\mu$m emission lines of our sources \citep{gruppioni16}. Assuming the Wide survey sensitivity and a signal-to-noise ratio threshold of 5, we found that a total observing time of $\sim24$hrs is sufficient to detect the 10 most luminous sources at $z>2$ in all the three emission lines. At $z>4$, detecting the 10 most luminous AGN with both [Ne$\,\textsc{v}$]$14.3\mu$m and [O$\,\textsc{iv}$]$25.9\mu$m requires $\sim90$ hours of integration time. Finally, considering only [O$\,\textsc{iv}$] detections, the 100 most luminous AGN at $z>2$ can be detected in $\sim150$hrs. For a more detailed analysis of the use of AGN- and star formation-related emission lines in characterizing AGN and host-galaxy properties, we refer to B24.

Besides AGN-related lines, we also expect to detect the SF-related Polycyclic Aromatic Hydrocarbons (PAHs) features \citep{leger89}, as well as fine structure lines of O, C, Ne, S, N, Fe, Ar, and Si that can be used to probe the neutral and ionized gas and to estimate redshift, amount of gas, the contribution of AGN and SF to the continuum emission, and the presence of outflows \citep{meixner19, mordini21,bisigello24}.

\section{Conclusion}\label{sec:conclusions}
We investigated the capabilities of the \prima\ instrument \textsl{PRIMAger} in detecting AGN, with a particular emphasis on the synergies between \prima\ and \athena\ in identifying and characterising obscured AGN.\par
Using X-ray background synthesis models, we predicted the number of AGN as a function of redshift, luminosity, and obscuration, and simulated the fraction of sources detectable by \athena\ in Deep ($1\,\rm{deg^2}$) and Wide ($28\,\rm{deg^2}$) surveys. For each redshift, luminosity, and obscuration bin, we assigned a set of SEDs from real AGN in the COSMOS field. By convolving the fluxes of these SEDs with the \textsl{PRIMAger} instrumental response, we evaluated the detection capabilities of \prima. Specifically, we simulated two \prima\ surveys, each with a $1,000$hr exposure time and covering the same areas as the \athena\ surveys. Additionally, we compared our predictions to those achieved with the deepest \textit{Herschel} surveys covering similar regions. Finally, we simulated \prima\ measurements of the BHAD evolution, starting from a $\sim100\,\mu$m survey. Our main results are the following:

\begin{enumerate}
    \item The capabilities of \prima\ in measuring the BHAD will surpass those of any current survey. We anticipate achieving precise measurements of the BHAD up to $z\sim8$, enabling us to determine definitively whether the discrepancy between X-ray-derived estimates and theoretical predictions arises from a population of heavily obscured AGN. Furthermore, \prima\ will provide an unprecedented opportunity to study the evolution of AGN and galaxies, probing back to an age of the Universe of $\sim700\,$Myr. While the BHAD values we obtained depend on the assumed evolution of the total number of AGN, the associated uncertainties do not and effectively illustrate \prima\ potential in measuring the high-redshift evolution of the BHAD free from obscuration biases. Indeed, we showed that starting from IR-derived BHADs, we can effectively recover the BHAD with even greater accuracy.
    \item A Deep \prima\ survey will be extremely powerful in revealing AGN up to $\sim 4$ and beyond. We expect to detect $\sim30\%$ of all the AGN up to $z\sim10$ in all the 16 bands and more than $70\%$ in at least one band. On average, we expect to have 7 detections per (detected) source. We predict that we will be able to reveal almost the entire AGN population at the ``cosmic noon''. A Wide survey, on the other hand, will rarely detect AGN $z>2 $. However, thanks to its large area, we expect up to $3 \times 10^5$ AGN, of which $\sim30,000$ with complete detections across all 16 \textsl{PRIMAger} bands.
    \item The \athena\ capabilities in detecting AGN are heavily affected by the source obscuration. With the current survey strategy, most of the CT-AGN (except for the most luminous ones) will be completely missed. For less obscured sources, we will be able to detect most of them up to $z>2$. Independently from the source obscuration, the most luminous AGN will be visible up to $z\sim8-10$.
    \item We found the combination of \prima\ and \athena\ to be a powerful tool to completely sample the AGN population up to very high-$z$ regardless of the source obscuration. \athena\ will be able to detect the most luminous sources and almost all the unobscured AGN, while \prima will be very effective in recovering the heavily obscured and CT-AGN that \athena\ will miss. For these sources, it has been shown that SED-fitting or spectroscopic follow-ups with FIRESS will allow us to effectively recognise them as AGN and to characterise their properties.
    \item The synergies between the two instruments can also be exploited for the source characterisation.
    The large number of \prima\ filters will provide a substantial improvement in our capabilities to perform SED-fitting and properly characterise the host-galaxy and AGN properties at high-$z$. Moreover, the \athena-provided X-ray properties can be used to further constrain the SED fitting and overcome the AGN-SF degeneracies. Finally, at high-$z$ even just an X-ray detection should be enough to identify a source as an AGN. 
    \item Our comparison with the capabilities of existing \textit{Herschel} surveys shows that \prima\ will provide an exceptional improvement over existing FIR surveys, both in terms of numbers of sources and redshift. For the Deep survey, the fraction of detectable AGN at $98\mu$m increases from  $\sim10\%$ to $\sim60\%$, with most of them having multiple detections in the other \textsl{PRIMAger} bands. The Wide survey perfectly illustrated the capabilities of \prima\ in effectively performing large area surveys: we expect an increase in the number of $98\mu$m-detected sources by a factor $\sim30$ with respect to \textit{Herschel} surveys covering similar areas.
\end{enumerate}

In this work, we demonstrated the capabilities and the necessity of exploiting the synergies between \prima\ and \athena\ to detect the full AGN population at and beyond the ``cosmic noon'', and to effectively characterise these sources. We also highlighted the significant improvement \prima\ offers in detecting obscured AGN compared to current surveys and in accurately measuring the evolution of the BHAD. With no planned cryogenically-cooled FIR observatories apart from \prima, we emphasize that without this mission, the astronomical community will face a critical gap in FIR coverage through and beyond 2040. This absence would severely hinder our ability to study obscured AGN and dust-rich galaxies at high redshifts and to finally constrain the evolution of the BH accretion rate density.

\subsection*{Disclosures}
The authors have no relevant financial interests in the manuscript or other potential conflicts of interest.

\subsection* {Code, Data, and Materials Availability} 
The software POMPA utilised to estimate the AGN density is available at \hyperlink{http://www.bo.astro.it/~gilli/counts.html}{here}. The \athena\ responses and background can be accessed in \hyperlink{https://www.mpe.mpg.de/ATHENA-WFI/response_matrices.html}{link}. The SEDs utilised in this work were obtained from \citet{delvecchio15}, and can be provided upon request. The latest \prima\ capabilities can be found \hyperlink{https://prima.ipac.caltech.edu/page/instruments}{here}.
 
\begin{acknowledgements}
LB, LM, and MV acknowledge financial support from the Inter-University Institute for Data Intensive Astronomy (IDIA), a partnership of the University of Cape Town, the University of Pretoria and the University of the Western Cape, and from the South African Department of Science and Innovation’s National Research Foundation under the ISARP RADIOMAP Joint Research Scheme (DSI-NRF Grant Number 150551) and the CPRR HIPPO Project (DSI-NRF Grant Number SRUG22031677). FJC acknowledges funding from grant PID2021-122955OB-C41 funded by MCIN/AEI/10.13039/ 501100011033 and by ERDF A way of making Europe. Part of the research activities described in this paper were carried out with contribution of the Next Generation EU funds within the National Recovery and Resilience Plan (PNRR), Mission 4 - Education and Research, Component 2 - From Research to Business (M4C2), Investment Line 3.1 - Strengthening and creation of Research Infrastructures, Project IR0000034 – ``STILES - Strengthening the Italian Leadership in ELT and SKA''. ID acknowledges funding by the European Union – NextGenerationEU, RRF M4C2 1.1, Project 2022JZJBHM: ``AGN-sCAN: zooming-in on the AGN-galaxy connection since the cosmic noon'' - CUP C53D23001120006.
\end{acknowledgements}

\bibliographystyle{aa.bst} 
\bibliography{prima_newathena.bib} 

\end{document}